\documentstyle[preprint,psfig,aps]{revtex}
\begin{document}
\draft
\title{Oscillator strengths with pseudopotentials}
\author{K. Yabana\footnote{E-mail address yabana@carrot.sc.niigata-u.ac.jp}}
\address{Department of Physics, Niigata University\\
Niigata, Japan\\
and\\}
\author{G.F. Bertsch\footnote{E-mail bertsch@phys.washington.edu}}
\address{Physics Department and Institute for Nuclear
Theory\\
University of Washington, Seattle, WA 98195 USA
}
\maketitle
\def\C60{C$_{60}$}
\def\eq#1{eq. (\ref{#1})}
\begin{abstract}
The time-dependent local-density approximation (TDLDA) is
shown to remain accurate in describing the atomic response
of group IB elements under the additional approximation of 
using pseudopotentials to treat the effects of core electrons.
This extends the work of Zangwill and Soven who showed the utility of 
the all-electron TDLDA in the atomic response problem.
\end{abstract}
\def\be{\begin{equation}}
\def\ee{\end{equation}} 
\section{Introduction}  

Mean field theory is now widely used in chemistry and condensed-matter
physics, treating the electron-electron interaction in the local-density
approximation\cite{jo89}.    An additional approximation which
is often made is in multi-atom calculations is to use pseudopotentials
for the ionic part of the Hamiltonian\cite{ba82} to avoid explicit 
calculation of the core electrons. The static theory is remarkably successful 
in describing binding and ground state properties.
The dynamic theory based on
these Hamiltonian approximations, the time-dependent local-density
approximation (TDLDA) is also quite tractable numerically and often gives an
excellent description of the strong
transitions\cite{ru96,ya94,lu94,ya96,ya97,ja96,bl95}.

In simple systems such as atoms TDLDA computations are quite feasible
without making the pseudopotential approximation\cite{za80,pa84,st95,st97}. 
However for many-atom systems with three-dimensional bases such as
plane-wave or coordinate-space meshes, the use of pseudopotentials
is practically unavoidable.  This is the general motivation of
this study comparing the pseudopotential approximation to the
all-electron calculation of the atomic response. The quality of the 
TDLDA approximation depends of
course on the excitation energy and on the detail one requires.
At low energies, in the region of the discrete transitions, the
TDLDA provides overall account of the oscillator strength as function of
excitation, but the details of transition energies and strengths
are given much more accurately by more sophisticated techniques.
At energies above the ionization threshold the experimental spectra
do not have so much detail and the TDLDA is arguably
the most effective approximation available.  As one goes up in energy
the dynamic electron-electron interaction decreases in importance, 
and the TDLDA becomes an unnecessary refinement on the independent-electron
Hartree-Fock approximation.  Of course about the thresholds
for core excitation the pseudopotential TDLDA invalid, since
it doesn't treat core degrees of freedom explicitly.

\section{Oscillator strengths in IB atoms}
The question of the validity of the pseudopotential approximation in the
lower energy domain arose in our studies when we tried to apply TDLDA to
clusters made from group IB elements, i.e. Cu, Ag, and Au.  In these 
elements, the closed $d$ shell just below the valence $s$ shell is 
important to the dynamics and
cannot be treated in a frozen approximation.  However, we found when we
calculated the TDLDA response of small Ag clusters including the $d$-shell
electrons that the $f$ sum rule for the valance electrons was badly
violated. Since the $f$-sum rule counts the number of electrons, we expected
the sum rule to equal the number of valence electrons in the
pseudopotential calculation.  In fact, this is not the case because the
pseudopotential is nonlocal.  The pseudopotentials necessarily depend
on the angular momentum $l$ of the electron (with respect to the
ion), and the double commutator that gives the sum rule for a closed
$l$-shell has 
a contribution from the potential given by
\be
\Delta f_l = {2 l +1\over 3}{2 m\over \hbar^2} \int r^4\, dr |\phi_l(r)|^2
\left(-V_l(r)+ \sum_\pm(l\,0\,1\,0|l\pm1\,0)^2V_{l\pm1}\right) 
\ee
where $\phi_l(r)$ is the radial wave function of the electron
and $V_l(r)$ is the pseudopotential for angular momentum $l$.
In the IB atoms with explicit treatment of outer $s$- and $d$-electrons,
the ordinary sum rule arise from the kinetic energy operator gives $f=11$,
the number of active electrons.  The potential
roughly doubles this, which may be seen from the numbers in
Table I. The second column gives the total oscillator strength with
only the kinetic part of the Hamiltonian, and the third column gives
the result including the pseudopotential contribution from eq. (1)
as well.  The qualitative effect of a nonlocal pseudopotential on
oscillator strengths in clusters of the group-IA atom lithium has been 
recently discussed in ref. \cite{al97}.  Here
the nonlocality induces an effective mass to lower the oscillator
strength of the collective valence transition.

There are several
possibilities to interpret the large oscillator strength in the
pseudopotential of IB atoms. The best outcome for theory would
be if the pseudopotential TDLDA were still accurate in the low-energy
domain, and the extra $f$ strength is physical and due to indirect effects
of the core electrons. Another possibility is that the extra strength is an
artifact of the nonlocality of the pseudopotentials.  In that case we ask
further whether the approximation introduces spurious strength in the
spectroscopic domain or only in the higher energy domain, where it would be
less significant to present applications of the TDLDA.

We shall study the atomic response of group IB elements following
closely the method of Zangwill and Soven \cite{za80}.
The independent-electron response, given by 
\be
\Pi^0(\vec r,\vec r',\omega)=
\sum_{i,j} \phi_i^*(\vec r)\phi_i(\vec r')\phi_j^*(\vec r')\phi_j(\vec r)
{2 (e_i -e_j) \over (e_i -e_j)^2 -\omega^2 -i\eta},
\ee
is represented on a radial coordinate space mesh with an angular
momentum decomposition.  The sum over particle states in eq. (2)
is replaced by the single-particle Green's function, greatly simplifying the
treatment of the continuum\footnote{This technique
was first applied to calculate nuclear response
functions\cite{sh75}.}.  The interacting response is then computed
by the matrix equation,
\be
\Pi^{TDLDA} = \Pi^0 (1-v \Pi^0)^{-1}
\ee 
where $v$ is the electron-electron
interaction.  The interaction $v$ includes the local density approximation
to the exchange and correlation energy given by the parameterization
of Ceperley and Alder\cite{ce80}.  The Hamiltonian and Green's function
is nonrelativistic except for the Au calculation, where a relativistic
treatment is necessary to get reasonable agreement with spectroscopic
properties.  The construction of a relativistic Green's function is
discussed in refs. \cite{mo74,hy84}.

Our pseudopotentials are calculated by the procedure of Troullier
and Martins \cite{tr91}.  There is a single parameter
in constructing the pseudopotential, the radius $a$ at which the
potential joins the all-electron self-consistent potential, which is
used for the outer regions.
In this study we have used values $a = 1.1$ \AA for $s$, $d$, and $f$
orbitals in Cu and Ag, and $a=1.21$ \AA for the $p$ orbitals.  In Au, we
took $a=1.24$ \AA for all orbitals. As a consistency check, we show in Table I
the integrated TDLDA response calculated up to 400 eV.  The integrated
response $I(\omega)$ is given by the following integral over ${\rm Im}\, \Pi$,
\be
I(\omega) = \int_0^\omega\,d \omega' {d f\over d \omega}={2 m\over \pi
\hbar^2}
\int_0^\omega\,d \omega' \int\, dr dr'\, z z'\, {\rm Im}\, \Pi^{TDLDA}(r,r',\omega')
\ee
It may be seen from Table I that TDLDA
conserves the sum rule and the integrated strength agrees with
the double commutator, as it must.

We now compare the pseudopotential response with the all-electron response
in the different energy domains, first examining the spectroscopic 
transitions. The strong $s\rightarrow p$
excitation is calculated in various approximations with the results
shown in Table II.  In the case of Au, the energies and transition
strengths are the weighted averages for the $s_{1/2}\rightarrow p_{1/2}$
and $s_{1/2}\rightarrow p_{3/2}$ excitations.  
The independent-particle LDA is computed from the 
$\Pi^{0}$ response and is shown in the first column.  The transition
in the TDLDA, shown in the next column, has nearly the same energy
but a quenched strength due to screening by the $d$-shell electrons.
The screening effect amounts to a 40-60\% reduction of the transition
strengths.  This illustrates the advantage of the TDLDA that it incorporates
the screening automatically, unlike some other treatments\cite{ha78,mi78}.
The pseudopotential approximation gives 
very similar energies and transition strengths, as shown in the 
third column. For
completeness we also compare with experiment, although the 
well-known deficiencies of the LDA make this an unreliable 
application.  The empirical strengths show a screening somewhere
between the TDLDA prediction and the independent-particle value,
with the TDLDA giving a better account of the strength for Ag and Au.

We next turn to the continuum domain. Fig.~1-3 shows
the integrated response for the IB atoms for the energy domain
0 -100 eV.
The steep
rise in $I$ below 10 eV is due to the pair of discrete transitions
$s,d\rightarrow p$ given in Table II.  One can also see a small feature around 50-70
eV due to the transition from a deeply $p$ state to the partially occupied
valence $s$ state.  This transition is absent in the pseudopotential
calculation.  Note that the TDLDA strength is quenched with respect to the
independent-particle response by about 10-20\% even up to the higher energies. 
Comparing the pseudopotential and all-electron
calculations, we see that they are practically indistinguishable in 
the case of Cu and Ag.  They also track well in Au below 80 eV except 
for the energy of the lowest transition, as was noted in the previous
paragraph.

Unfortunately, there does not seem to exist experimental that one
can compare to.   Fig. 13 in ref. \cite{fa68} shows a 
curve for Cu, but it was measured for a thin film rather than the
isolated atom.

\section{Conclusion}

   We have found that the TDLDA with a pseudopotential 
approximation gives virtually the same distribution of oscillator 
strength in the
region 0 - 100 eV as the all-electron theory of IB atoms, despite the fact
that the $f$ sum rule is badly violated by the state-dependent
interaction.  We conclude that the pseudopotentials in these atoms 
properly take
into account many-body effects by the nonlocality of the potential,
and may be used confidence in studying the response of 
these elements.

We thank J. Fuhr for helpful information. This work is supported in part by 
the Department of Energy  under Grant DE-FG-06-90ER40561.

\begin{table}
\caption{Oscillator strengths $f$ for pseudopotential calculations of 
IB atoms}
\begin{tabular}{cccc}
         &          &    with         &   TDLDA\\
Element  &  kinetic & pseudopotential &   $I(400)$ \\
\tableline
Cu &   11   & 24.9 & 21.5\\
Ag &   11   &  19.5 & 19.4\\
Au &  11    &  21.9 & 21.9 \\    
\end{tabular}
\end{table}

\begin{table}
\caption{Energies and strengths of the
$s\rightarrow p$ transition in IB metal atoms}
\begin{tabular}{ccccccc}
& &free & TDLDA & TDLDA &many-body& exp. \\
& &     & all-electron & pseudopotential & \cite{jo90}&\cite{re80,fu95}\\   
\tableline
Cu &  $E$ (eV)& 4.1&4.3&4.3&3.9&3.8\\
& $f$ & 0.9 &0.4 & 0.4 && 0.66 \\
\tableline 
Ag &  $E$ (eV)& 3.6 &3.9 &4.2&& 3.7 \\
                  & $f$ & 1.0 & 0.6 & 0.6 && 0.7 \\
\tableline
Au &  $E$ (eV) & 5.1 & 5.2 & 5.3 & & 4.9\\
   & $f$       & 1.1 & 0.33 & 0.38 &&  0.5\\
\end{tabular}
\end{table}
\section{Figure Captions}
\noindent
Fig.~1  Integrated transition strength in Cu:  all-atom TDLDA
(solid line); pseudopotential TDLDA (short dashed line);
static LDA (long dashed line).\\
\\
Fig.~2  Integrated transition strength in Ag:  all-atom TDLDA
(solid line); pseudopotential TDLDA (short dashed line);
static LDA (long dashed line).\\
\\
Fig.~3  Integrated transition strength in Au:  all-atom TDLDA
(solid line); pseudopotential TDLDA (short dashed line);
static LDA (long dashed line).

\begin{figure}
  \begin{center}
    \leavevmode
    \parbox{0.9\textwidth}
           {\psfig{file=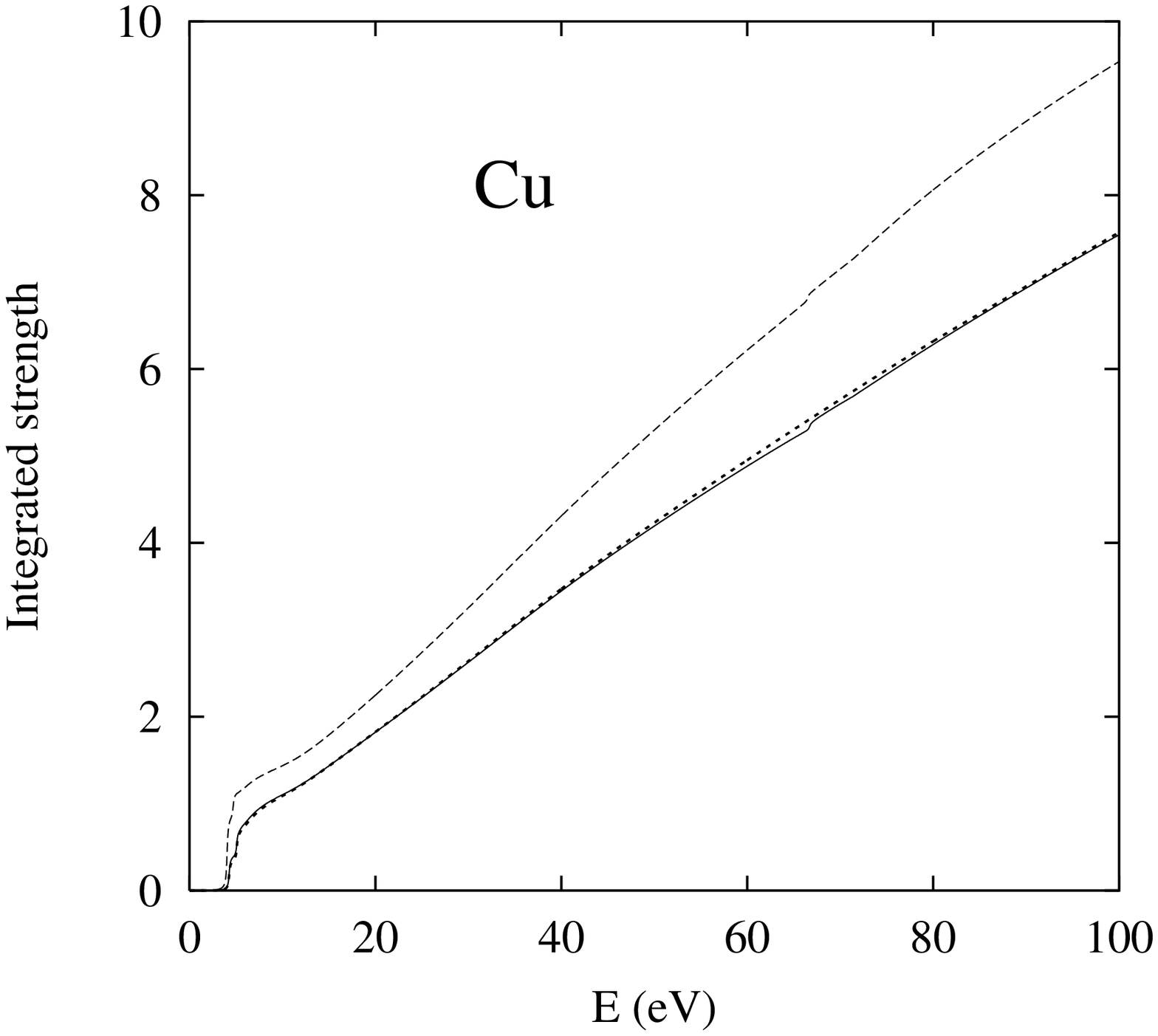,width=0.9\textwidth}}
  \end{center}
\protect\caption{
Integrated transition strength in Cu:  all-atom TDLDA
(solid line); pseudopotential TDLDA (short dashed line);
static LDA (long dashed line).
}
\end{figure}
\begin{figure}
  \begin{center}
    \leavevmode
    \parbox{0.9\textwidth}
           {\psfig{file=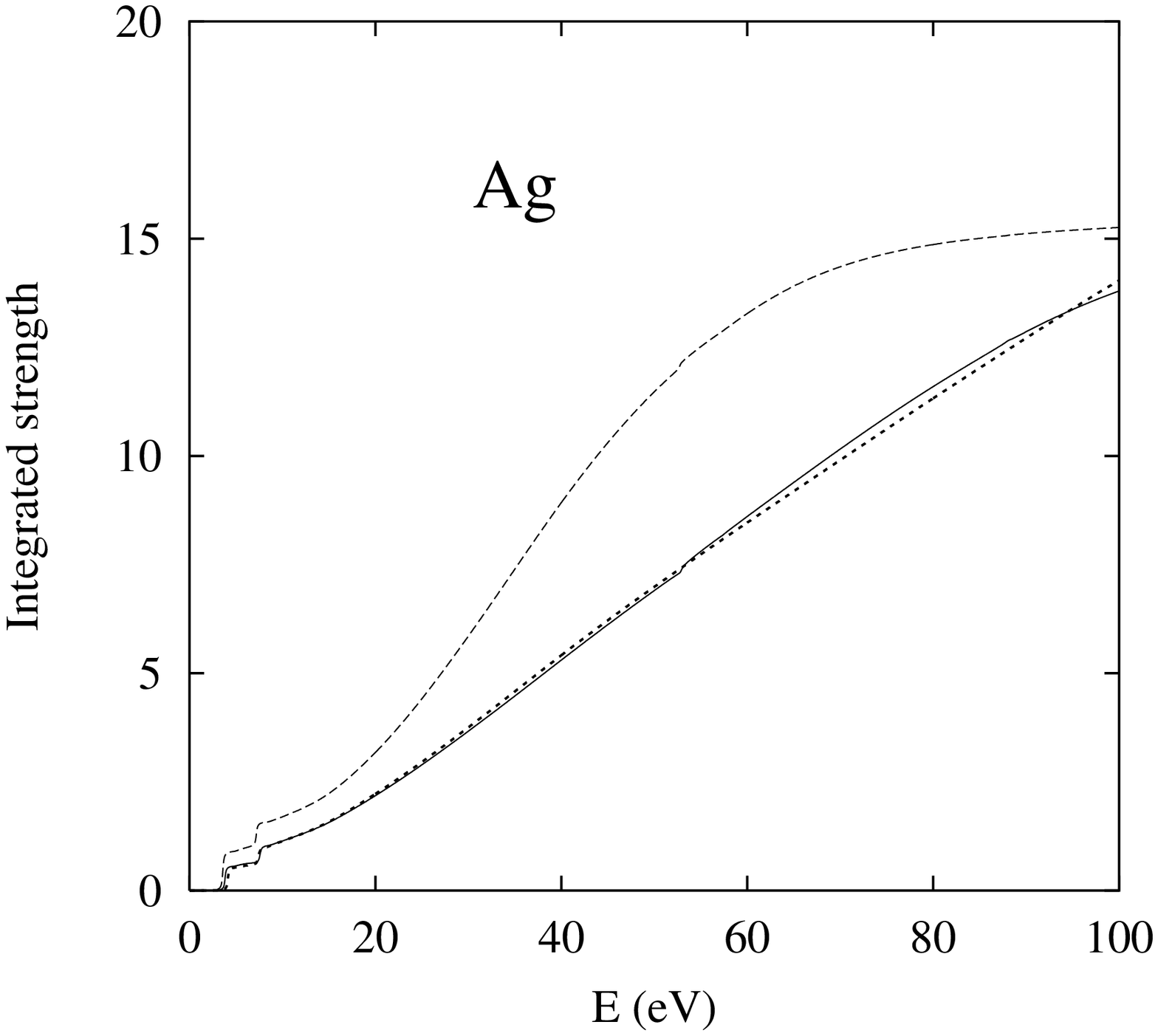,width=0.9\textwidth}}
  \end{center}
\protect\caption{
Integrated transition strength in Ag:  all-atom TDLDA
(solid line); pseudopotential TDLDA (short dashed line);
static LDA (long dashed line).
}
\end{figure}

\begin{figure}
  \begin{center}
    \leavevmode
    \parbox{0.9\textwidth}
           {\psfig{file=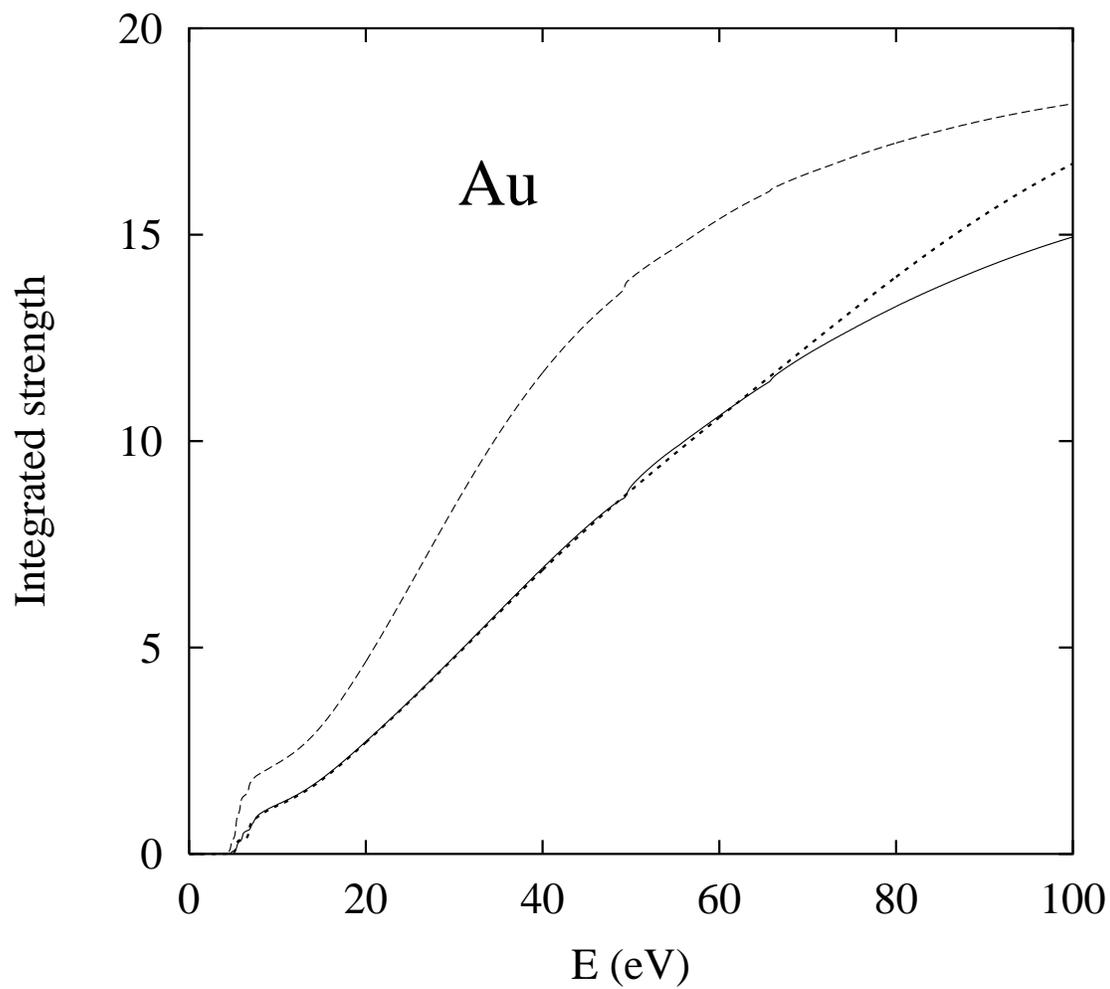,width=0.9\textwidth}}
  \end{center}
\protect\caption{
Integrated transition strength in Au:  relativistic all-atom TDLDA
(solid line); pseudopotential TDLDA (short dashed line);
relativistic static LDA (long dashed line).
}
\label{fig1}
\end{figure}

\end{document}